\documentclass[twocolumn,preprintnumbers,amsmath,amssymb,superscriptaddress,footnote]{revtex4-2}
\usepackage{graphicx} 
\usepackage[dvipdfmx]{epsfig} 
\usepackage{bm}
\usepackage{ulem}
\usepackage{amsmath,color}

\def\m@thcombine#1#2{
  \setbox0=\hbox{$#1$}
  \setbox1=\hbox{$#2$}
  \ifdim\wd0>\wd1
    \setbox0=\hbox to\wd1{\hss\box0\hss}
  \else
    \setbox1=\hbox to\wd0{\hss\box1\hss}
  \fi
  \mathop{\vcenter{
    \offinterlineskip\box0\box1}}}
\def\lesim{\m@thcombine<\sim}
\def\gesim{\m@thcombine>\sim}

\begin{document}

\title{Evidence of bicluster structure in the ground state of $^{20}$Ne}
\author{Y. Yamaguchi}
\affiliation{Department of Physics, Osaka Metropolitan University, Osaka 558-8585, Japan}
\author{W. Horiuchi}
\email{whoriuchi@omu.ac.jp}
\affiliation{Department of Physics, Osaka Metropolitan University, Osaka 558-8585, Japan}
\affiliation{Nambu Yoichiro Institute of Theoretical and Experimental Physics (NITEP), Osaka Metropolitan University, Osaka 558-8585, Japan}
\affiliation{RIKEN Nishina Center, Wako 351-0198, Japan}
\affiliation{Department of Physics,
  Hokkaido University, Sapporo 060-0810, Japan}

\author{N. Itagaki}
\email{itagaki@omu.ac.jp}
\affiliation{Department of Physics, Osaka Metropolitan University, Osaka 558-8585, Japan}
\affiliation{Nambu Yoichiro Institute of Theoretical and Experimental Physics (NITEP), Osaka Metropolitan University, Osaka 558-8585, Japan}
\affiliation{RIKEN Nishina Center, Wako 351-0198, Japan}

\begin{abstract}
  We explore the structure of the ground state of $^{20}$Ne by investigating
  various density profiles.
  Four candidates for the ground state configurations,
  (a) $j$-$j$ coupling and (b) SU(3) shell model and
  (c) $5\alpha$ and (d) $^{16}{\rm O}+\alpha$ cluster model configurations
  are generated by utilizing the antisymmetrized quasicluster model.
  A high-energy reaction theory, the Glauber model,
  relates these one-body density distributions and reaction observables.
  The angular distributions of
  the elastic scattering cross sections clearly distinguish
  these configurations and tell which is the most plausible one: 
  The ground state of $^{20}$Ne favors a 16+4 nucleon bicluster structure.
  A comprehensive investigation of other electric observables
  also supports this conclusion.
\end{abstract}
\maketitle

\preprint{NITEP 174}

\section{Introduction}

Nuclear clustering phenomena often appear
in light $N=Z$ nuclei~\cite{Ikeda68, Ikeda80,Fujiwara80,Freer18}.
Accounting for the $\alpha$ ($^{4}$He) clustering
is essential to understand their low-lying states.
Especially, the first excited $J^\pi=0^+$ states 
play an important role in explaining nucleosynthesis
producing $^{12}$C and $^{16}$O elements~\cite{Hoyle54,deBoer17}
and they are well explained by $3\alpha$~\cite{Kamimura81}
and $^{12}{\rm C}+\alpha$~\cite{Suzuki76a,Suzuki76b} cluster models,
respectively.
See Ref.~\cite{Fujiwara80} for
a comprehensive review of these $\alpha$ clustered nuclei.
Clustering phenomena are closely related to the bosonic property
of the nuclear system. Ref.~\cite{Tohsaki01} proposed that
some multi-$\alpha$ states can be interpreted as
a Bose-Einstein condensed state.
Experimental searches for the $\alpha$ condensed states
have been made for $^{16}$O~\cite{Wakasa07} and $^{20}$Ne~\cite{Adachi21}.

Quantifying the degree of nuclear clustering
is one of the hot topics in nuclear physics.
An analysis of the $\alpha$ knockout reaction for $^{20}$Ne
was performed, and the degree of the $^{16}{\rm O}+\alpha$ cluster structure
was quantified~\cite{Yoshida19}. The $\alpha$ clustering is also found in
the surface region of heavy nuclei such as Sn isotopes~\cite{Tanaka21}.
Recently, two of the present authors (W.H. and N.I.) developed
an efficient way to visualize the ground state structure by using
the proton-nucleus elastic scattering combined
with the antisymmetrized quasicluster
model (AQCM)~\cite{Horiuchi22c,Horiuchi23}.
The shell and cluster configurations can easily be distinguished
in comparison with angular distributions of the cross sections at the
first diffraction peak.

It should be noted that the ground state structure of $^{20}$Ne is still controversial and we need comprehensive 
understanding. The structure of $^{20}$Ne
has been recognized as having $^{16}{\rm O}+\alpha$ structure,
which explains positive and negative parity rotational bands,
the so-called inversion doublet~\cite{Horiuchi68,Fujiwara80}.
The $5\alpha$ cluster model calculation showed that
the most probable ground state structure is
the $^{16}{\rm O}+\alpha$ bicluster structure~\cite{Zhou14}.
On the other hand, the non-clustered rotational model
can also reproduce the electric properties
of the low-lying states of $^{20}$Ne~\cite{Horikawa71,Horikawa72,Hatakeyama19}.
A fully microscopic description of $^{20}$Ne suggested that the coexistence
of deformed mean-field and cluster pictures in the low-lying spectrum of
$^{20}$Ne~\cite{Abgrall74,Kanada95,Kimura04,Marevic18}. 

Furthermore, in Ref.~\cite{Horiuchi23},
the AQCM analysis was applied to $^{16}$O and concluded that
the structure of the ground state has most likely
4$\alpha$ tetrahedron configuration
in line with modern $ab\ initio$ calculations~Ref.~\cite{Epelbaum14}.
One may ask a question: If this interpretation is accepted,
the ground-state structure of $^{20}$Ne could have $5\alpha$ structure,
while $^{20}$Ne structure has been recognized as
the $p$-shell closed $^{16}$O plus $\alpha$ structure.
There is a need to clarify the most probable structure of $^{20}$Ne
using different observables.

In this paper, we perform the AQCM analysis to unveil
the ground state structure of $^{20}$Ne.
The density profiles generated from different AQCM configurations
are investigated by using proton-nucleus elastic scattering.
We also compute other observables and compare them with experimental data
to see more details about the density distributions.

In the next section, we explain the theoretical framework
used in this study. The $^{20}$Ne wave function with the AQCM is defined.
The evaluation of the one-body density distribution is explained,
which is used as input to the high-energy reaction theory, the Glauber model.
Section~\ref{results.sec} explains how we generate
various AQCM configurations in detail. 
We discuss the properties and density profiles
of the four specific configurations.
Using these one-body density distributions,
the proton-$^{20}$Ne elastic scattering cross sections
and elastic charge form factors are computed
and compared with available experimental data.
The conclusion is made in Sec.~\ref{conclusion.sec}.

\section{Method}
\label{method.sec}

\subsection{Configurations of $^{20}$Ne with antisymmetrized quasicluster model}

The configurations of $^{20}$Ne are generated by the antisymmetrized quasicluster model (AQCM)~\cite{AQCM01,AQCM02,AQCM03,AQCM04,AQCM05,AQCM06,AQCM07,AQCM08,AQCM09,AQCM10,AQCM11,AQCM12,AQCM13,AQCM14}.
The AQCM is an extension of the multi-$\alpha$ cluster model~\cite{Brink},
which allows one to express not only cluster model
but also the $j$-$j$ coupling and SU(3) shell model configurations
in a single scheme.
The AQCM basis state for $^{20}$Ne is expressed
by the $5\alpha$ cluster model as
\begin{align}
  \Phi(\nu,d,d_5,\Lambda)=\mathcal{A}
  \left\{\prod_{i=1}^5 \Phi_\alpha(\nu,R_i,\Lambda)\right\},
\end{align}
where $\mathcal{A}$ is the antisymmetrizer for nucleons.
Here $\Phi_{\alpha}(\nu,R_i,\Lambda)$ is
the $i$th $\alpha$ particle wave functions
with the Gaussian center parameter $\bm{R}_i$ defined by
\begin{align}
  \Phi_{\alpha}(\nu,R_i,\Lambda)=
  \phi^\nu_1(\uparrow,p)\phi^\nu_2(\downarrow,p)\phi^\nu_3(\uparrow,n)\phi^\nu_4(\downarrow,n)
\end{align}
with a single-nucleon Gaussian wave packet
with spin $\chi_s$ ($s=\uparrow$ or $\downarrow$)
and isospin $\eta_t$ ($t=p$ or $n$) wave functions
\begin{align}
  \phi^\nu_j(s,t)=  \left(\frac{2\nu}{\pi}\right)^{3/4}\exp\left[-\nu(\bm{r}_j-\bm{\zeta}_i)^2\right]\chi_{s}\eta_{t},
\end{align}
where
\begin{align}
  \bm{\zeta}_i=\bm{R}_i+i\Lambda \bm{e}^{\rm spin}\times \bm{R}_i
\end{align}
with $\bm{e}^{\rm spin}$ being a unit vector for the intrinsic-spin
orientation of a nucleon.
Note that the AQCM basis state
  with $\Lambda=0$ is nothing
  but the Brink model~\cite{Brink}, 
  where the spin-saturated $\alpha$ cluster is assumed and
  no contribution from the spin-orbit interaction is realized.
  To express the independent particle motion
  with a good total angular momentum $j$, the imaginary part of
  the Gaussian center parameter is introduced in the AQCM.

The basis function of $^{20}$Ne is expressed by the four-$\alpha$ part and
the additional 5th $\alpha$ wave function.
In the four-$\alpha$ part, the four $\alpha$ particles 
are placed at vertexes of a tetrahedron
with a common inter-$\alpha$-cluster
distance $d=|\bm{R}_i-\bm{R}_j|$ ($i\neq j \leq 4$).
Note that one can obtain the $p$-shell
closure configuration $(0s)^4(0p)^{12}$ for $^{16}$O
with a limit of $d\to 0$~\cite{Brink}
independent of a choice of $\Lambda$.
Here we place the 5th $\alpha$ particle along the $z$ axis passing through
the center of the bottom face of the tetrahedron.
The distance between the 5th $\alpha$ particle
and the center of the tetrahedron is denoted as $d_5$.
With this geometry of the $\alpha$ particles,
one can obtain both the SU(3) [$(0s)^4(0p)^{12}(1s0d)^{4}$]
and the $j$-$j$ coupling [$(0s)^4(0p)^{12}(0d_{5/2})^{4}$]
shell model configurations by taking $\Lambda=0$ and 1, respectively,
in the limit of $d, d_5\to 0$~\cite{AQCM04}.

Once the parameters of the AQCM basis function, i.e.,
$\nu$, $d$, $d_5$, and $\Lambda$, are fixed,
we can compute the nucleon one-body density distribution in
the body-fixed frame
with the mass number $A$ as
\begin{align}
  \tilde{\rho}(\bm{r})
  = \langle \Phi |\sum_{i=1}^A \delta(\bm{r}_i-\bm{r}) | \Phi \rangle / \langle \Phi | \Phi \rangle,
\end{align}
Note $\sum_{i=1}^A\left<\Phi\right|\bm{r}_i\left|\Phi\right>=0$.
$\tilde{\rho}$ in general includes the center-of-mass motion.
This is properly removed by using the prescription
given by Ref.~\cite{Horiuchi07}
\begin{align}
  \int d\bm{r}\,e^{i\bm{k}\cdot\bm{r}}\rho_{\rm int}(\bm{r})=\exp\left(\frac{k^2}{8A\nu}\right)
  \int d\bm{r}\,e^{i\bm{k}\cdot\bm{r}} \tilde{\rho}(\bm{r}).
\end{align}
The density distribution in the laboratory frame is obtained
by averaging the center-of-mass-free density distribution
in the body-fixed frame over angles as
~\cite{Horiuchi12}
\begin{align}
  \rho(r)=\frac{1}{4\pi}\,\int d\hat{\bm{r}}\,\rho_{\rm int}(\bm{r}).
\label{density.eq}
\end{align}

\subsection{High-energy reaction observables}

To relate the density profile to the reaction observables,
we calculate the elastic scattering and total reaction cross sections using
a high-energy microscopic reaction theory, the Glauber model~\cite{Glauber}.
Here, we briefly explain how to get the cross sections 
with the one-body density distribution obtained above.

The scattering amplitude
of the proton-nucleus elastic scattering is given by~\cite{Suzuki03}
\begin{align}
  f(\theta)=F_C(\theta)+\frac{ik}{2\pi}\int d\bm{b}\, e^{-i\bm{q}\cdot\bm{b}+2i\eta \ln(kb)}\left(1-e^{i\chi_{pT}(\bm{b})}\right),
\end{align}
where $F_C(\theta)$ is the Rutherford scattering amplitude,
$\bm{b}$ is the impact parameter vector,
and $\eta$ is the Sommerfeld parameter.
The relativistic kinematics is used for the wave number $k$.
With this scattering amplitude,
the proton-nucleus elastic scattering differential cross section is
computed by
\begin{align}
  \frac{d\sigma}{d\Omega}=|f(\theta)|^2.
\end{align}

The optical phase-shift function $\chi_{pT}$ in
the optical-limit approximation (OLA) is given by~\cite{Glauber,Suzuki03}
\begin{align}
  i\chi_{pT}(\bm{b})\approx -\int d\bm{r}\,
  \left[\rho_p(\bm{r})\Gamma_{pp}(\bm{b}-\bm{s})
    +\rho_n(\bm{r})\Gamma_{pn}(\bm{b}-\bm{s})\right],
\end{align}
where $\bm{r}=(\bm{s},z)$ is
  the single-nucleon coordinate measured from the center of mass
  coordinate of the projectile nucleus
  with $z$ being the beam direction.
The inputs to the theory are the density
distribution obtained by Eq.~(\ref{density.eq})
and proton-proton (proton-neutron) profile function
$\Gamma_{pp}$ ($\Gamma_{pn}$), which is often parametrized as~\cite{Ray79}
\begin{align}
  \Gamma_{NN}(\bm{b})=\frac{1-i\alpha_{NN}}{4\pi \beta_{NN}}\sigma_{NN}^{\rm tot}
  e^{-\frac{b^2}{2\beta_{NN}}}
\end{align}
for each incident energy. The parameter sets of $\alpha_{NN}$, $\beta_{NN}$,
and $\sigma_{NN}^{\rm tot}$ are given
in Ref.~\cite{Ibrahim08}, which are well tested
for proton-nucleus scattering~\cite{Ibrahim09,Horiuchi16, Hatakeyama19, Horiuchi22c,Horiuchi23}.
As discussed in Ref.~\cite{Hatakeyama18},
the proton-elastic scattering has a strong sensitivity to the density
profiles near the nuclear surface and can be used
as a spectroscopic tool
to distinguish shell and cluster configurations in light- to medium-mass
nuclei~\cite{Horiuchi22c,Horiuchi23}.

The total reaction cross sections offer
direct observables of nuclear size properties
and are used to verify our wave function in the present study.
In the Glauber model~\cite{Glauber}, $\sigma_R$ is calculated by
\begin{align}
  \sigma_R=\int d\bm{b}\, \left(1-|e^{i\chi_{PT}(\bm{b})}|^2\right).
\end{align}
Here we employ the nucleon-target formalism in the Glauber model
(NTG)~\cite{NTG} to evaluate projectile-target optical phase-shift
function $\chi_{PT}(\bm{b})$, which only requires the one-body
density distributions of the projectile and target nuclei
and the profile function. We investigate $\sigma_R$
on a carbon target, in which the experimental data are available.
The harmonic-oscillator-type density that reproduces the charge radius
is employed for the carbon target.
It is known that a carbon target
has strong sensitivity to the density profiles
near the nuclear surface~\cite{Horiuchi14,Makiguchi22}.
We remark that the present reaction model has been used
as a standard tool for extracting nuclear size properties
from the interaction cross section measurements~\cite{Kanungo10,Kanungo11,Bagchi20}
as it works well shown in many examples of the nucleus-nucleus
scattering~\cite{Horiuchi06,Horiuchi07,Ibrahim09,Horiuchi10,Horiuchi12,Horiuchi15,Nagahisa18}. 

\section{Results}
\label{results.sec}

\subsection{Properties of $^{20}$Ne configurations}

  \begin{figure}[ht]
\begin{center}
  \epsfig{file=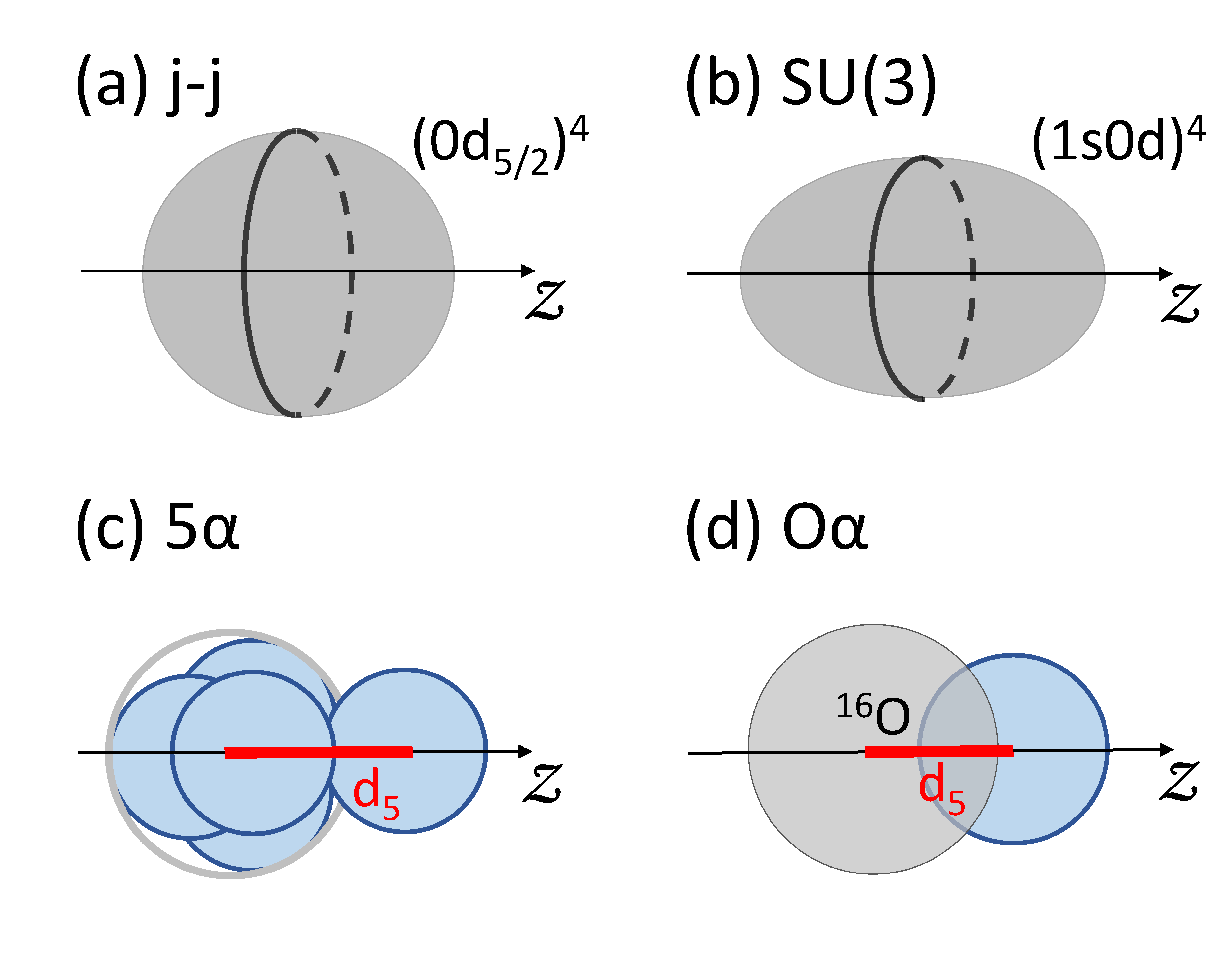, width=\linewidth}
  \caption{Schematic illustration of adopted configurations of $^{20}$Ne.
    (a) $j$-$j$ coupling and (b) SU(3) shell model configurations
    and (c) $5\alpha$ and (d) bicluster (O$\alpha$) configurations.
    Sizes of $\alpha$ and $^{16}$O clusters
      and their distance are drawn, reflecting
      the actual scales given in Table~\ref{results.tab}.}
    \label{conf.fig}
  \end{center}
\end{figure}

In this paper, we examine four types of density profiles
of $^{20}$Ne obtained from the AQCM basis functions with
proper choices of $\nu$, $d$, $d_5$ and $\Lambda$, extending
the direction made in Refs.~\cite{Horiuchi22c,Horiuchi23}.
We set all these generated AQCM configurations
to reproduce the measured charge radius
of $^{20}$Ne~\cite{Angeli13}.
The schematic pictures of these four configurations
are illustrated in Fig.~\ref{conf.fig}
and their details are described below.

First, we consider two types of shell-model configurations,
$j$-$j$ coupling and SU(3) shell models,
illustrated in panels (a) and (b) of Fig.~\ref{conf.fig}.
The $j$-$j$ and SU(3) configurations correspond
to $(0d_{5/2})^4$ and $(1s0d)^4$ configurations, respectively.
In the AQCM, the $j$-$j$ (SU(3)) configuration can be expressed
taking $\Lambda=1$ ($\Lambda=0$) in a limit of $d,d_5\to 0$.
The remaining oscillator parameter $\nu$ is fixed to reproduce the measured
charge radius of $^{20}$Ne~\cite{Angeli13}.
Hereafter they are called $j$-$j$ and SU(3) configurations, respectively.
The latter corresponds to the axially symmetric harmonic oscillator
model, which exhibits prolate deformation.

Next, two types of cluster-model configurations are prepared.
The $5\alpha$ configuration is generated by taking into account
the $4\alpha$ tetrahedron structure of $^{16}$O~\cite{Horiuchi23}.
As illustrated in panel (c), in the AQCM, a $5\alpha$-like cluster configuration
is expressed by taking the $\nu$ parameter of the free $\alpha$ particle
and the $d$ value reproducing the charge radius of $^{16}$O
with $\Lambda=0$~\cite{Horiuchi23}.
Then the distance of the 5th $\alpha$ particle, $d_5$,
is set to reproduce the measured charge radius of $^{20}$Ne.
This is called $5\alpha$ configuration.

Another cluster configuration represents a bicluster or
$^{16}$O+$\alpha$ like clustering,
which divides 20 nucleons into 16+4 nucleons drawn in panel (d)
of Fig.~\ref{conf.fig}.
The $4\alpha$ part is constructed based on the $p$-shell closed
shell model configuration, which
can be realized in the AQCM
by taking $d\to 0$ with $\nu$ reproducing 
the charge radius of $^{16}$O~\cite{Horiuchi23},
and then the $d_5$ value is fixed
to reproduce the measured charge radius of $^{20}$Ne.
Hereafter this is called O$\alpha$ or bicluster configuration.

Table~\ref{results.tab} lists the thus-obtained AQCM parameter sets,
and calculated nuclear properties with these four configurations of $^{20}$Ne:
the total harmonic oscillator quanta $\left<Q\right>$,
the expectation values of single-particle spin-orbit operators
$\sum_{i=1}^A\bm{l}_i\cdot\bm{s}_i$, $\left<LS\right>$,
and the single-particle parity operators
$\sum_{i=1}^AP_{i}$ with $P_if(\bm{r}_i)=f(-\bm{r}_i)$, $\left<P\right>$.
We confirm that the $\left<Q\right>$, $\left<LS\right>$, and
$\left<P\right>$ values of the shell model configurations reproduce
the expected values from the $j$-$j$ and SU(3) shell model configurations
given in parentheses.

Both the cluster configurations, $5\alpha$ and O$\alpha$ types, exhibit large $d_5$ values, about 3--4 fm,
indicating a well-developed cluster structure:
The $d$ and $d_5$ values are comparable in the $5\alpha$ configuration,
representing a trigonal bipyramid configuration.
In the O$\alpha$ configuration, the large $d_5$ value of 3.89 fm is found,
which is almost equal to the sum of the matter radii of the
$p$-shell closed $^{16}$O and four-nucleon cluster,
2.57 and 1.85~fm, respectively, 
indicating well developed 16+4 bicluster structure.

The $5\alpha$ configuration predict the largest $\left<Q\right>$ value,
while the O$\alpha$ configuration shows slightly larger $\left<Q\right>$ value
compared to the ideal shell model configuration, $\left<Q\right>=20$.
This is because, in the 5$\alpha$ configuration,
the $4\alpha$ configuration in $^{16}$O core already
includes high oscillator quanta as $\left<Q\right>=18.6$
originating from the large intercluster distances when
the $\nu$ value is large~\cite{Horiuchi23}.
This happens in general when a nuclear system exhibits
  well-developed cluster structure~\cite{Suzuki76a,Suzuki76b,Suzuki96,Neff12,Horiuchi14b,Horiuchi14c}.
As the charge radius of $^{20}$Ne is larger than
that of $^{16}$O, the additional 5th $\alpha$ particle
should be located at the surface
region of $^{16}$O, resulting in high $\left<Q\right>$ value.
In contrast, for the O$\alpha$ configuration,
since the $\nu$ value is much smaller than that of the free $\alpha$ particle,
the mixing of higher major shell components is suppressed.

\begin{table}[htb]
  \begin{center}
    \caption{Properties of the four specific AQCM configurations of $^{20}$Ne.
      Values in parentheses are obtained
      with the ideal shell-model configurations.
      See text for details.
      All the root-mean-square point-proton radii of these configurations 
      are commonly set to 2.89~fm~\cite{Angeli13}.}
\begin{tabular}{cccccccc}
  \hline\hline
  &$\nu$ (fm$^{-2}$) &$d$ (fm)& $d_5$ (fm)&$\Lambda$&$\left<Q\right>$ &$\left<LS\right>$&$\left<P\right>$ \\
\hline
$j$-$j$&0.1453&0.001&0.01&1&20.0(20)&3.99(4) &$-4.0$($-4$)\\
SU(3) &0.1453&0.001&0.01&0&20.0(20)&0.0(0)  &$-4.0$($-4$)\\
5$\alpha$  &0.2656&3.058&3.895&0& 32.3 &0.0(0)&$-1.5$\\  
O$\alpha$  &0.1635&0.001&3.012&0& 21.3 &0.0(0)&$-3.7$\\
\hline\hline
\end{tabular}  
\label{results.tab}
\end{center}
\end{table}

\begin{figure}[ht]
\begin{center}
  \epsfig{file=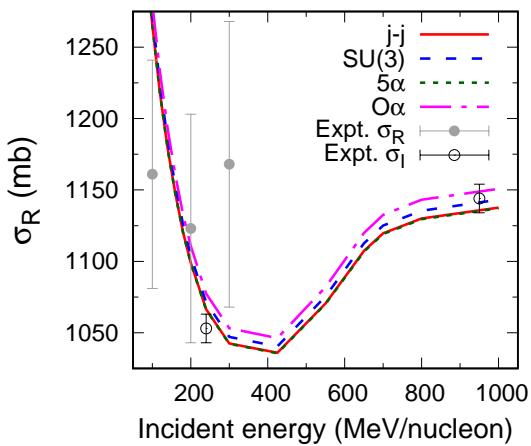, scale=1.3}
  \caption{Total reaction cross sections on a carbon target of $^{20}$Ne
    as a function of incident energy. The experimental total reaction $(\sigma_R)$
    and interaction cross section ($\sigma_I$) data
    are taken from Refs.~\cite{Kox86,Chulkov96,Takechi12}.}
    \label{rcs.fig}
  \end{center}
\end{figure}

To verify those four configurations, we calculate the
total reaction cross sections on a carbon target
using the density distributions obtained from these four configurations.
Figure~\ref{rcs.fig} plots
the calculated total reaction cross sections on a carbon target
as a function of incident energy.
The experimental total reaction ($\sigma_R$)
and interaction ($\sigma_I$) cross section data
are also plotted for comparison.
The uncertainties of the $\sigma_R$ data are quite large
and do not help constrain the theoretical results in the present paper.
The calculated $\sigma_R$ results reasonably reproduce
the experimental $\sigma_I$ data, considering
that the difference between $\sigma_R$ and $\sigma_I$
is tens of mb~\cite{Kohama08}
and the difference becomes larger
in general, as the incident energy decreases~\cite{Hatakeyama19}.
Noting that $\sigma_R>\sigma_I$ always holds,
the O$\alpha$ configurations give the best reproduction of the data.
However, the experimental uncertainties of $\sigma_I$ are comparable to
the differences in the total reaction cross sections
calculated with the four density profiles.
Other observables should be investigated
to conclude which configurations are the most probable in $^{20}$Ne.

\subsection{Density profiles of $^{20}$Ne}

\begin{figure}[ht]
\begin{center}
  \epsfig{file=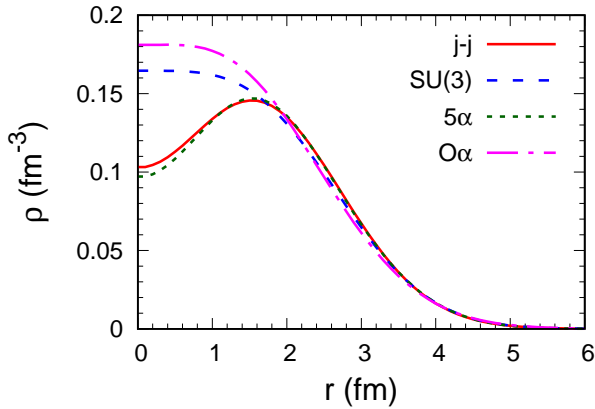, scale=1.3}
  \caption{Nucleon density distributions of $^{20}$Ne
  with the four types of configurations. See text for details.}
    \label{dens.fig}
  \end{center}
\end{figure}

The characteristics of these four configurations can be seen
in the density profiles.
Figure~\ref{dens.fig} plots the one-body density distributions of $^{20}$Ne.
The $j$-$j$ and $5\alpha$ configurations exhibit
the so-called ``bubble'' structure,
while the central depressions are filled
in the SU(3) and O$\alpha$ configurations.
The nature of the bubble structure is
the lack of occupation of the $s$ orbit.
The bubble structure of the $j$-$j$ configuration
is simply understood by $(0s)^4(0p)^{12}(0d_{5/2})^4$
  configuration, where the $1s$ orbit is absent.
For the $5\alpha$ configuration, the bubble structure is developed
as the additional $\alpha$ particle is located at the surface
of the $^{16}$O cluster and the $4\alpha$ configurations
of $^{16}$O part already have the bubble structure~\cite{Horiuchi23}.
For the SU(3) and O$\alpha$-cluster configurations,
the 5th $\alpha$ cluster should occupy the $sd$ shell
and the central depression disappears by the occupation of the $s$ orbit.
This can also be interpreted as the deformation effect
which induces the configuration mixing of
the single-particle orbits near the Fermi level~\cite{Horiuchi21c}.
We remark that the recent no-core shell model calculation
  predicts the $(\lambda,\mu)=(N_z-N_x, N_x-N_y)=(8,0)$ dominance
  in the ground state of $^{20}$Ne~\cite{Dytrych20},
  where $N_k$ is the oscillator quanta of the $k$ ($=x,y,z$)-direction.
  This $(\lambda,\mu)=(8,0)$ configuration
  corresponds the SU(3) configuration
  in the present paper, showing non-bubble structure.
  Note that the O$\alpha$ configuration is expressed
  by a superposition of $(\lambda,0)$ configurations 
  whose occupation numbers depend on the intercluster distance.
  As we see later, the the O$\alpha$ configuration includes a
  large component of the SU(3) configuration.

In most cases, the nuclear deformation diffuses the density profiles near
the nuclear surface compared to the spherical limit.
It is convenient to quantify the density profile near the nuclear surface.
For this purpose, we evaluate the diffuseness parameter
of the one-body density distribution using the prescription given
in Ref.~\cite{Hatakeyama18}, where
a two-parameter Fermi (2pF) function 
\begin{align}
\rho_{\rm 2pF}(\bar{R},a,r)=\frac{\rho_{0}}{1+\exp[(r-\bar{R})/a]},
\end{align}
is determined by the least square method by minimizing 
\begin{align}
  \int_{0}^\infty dr\,r^2\left|\rho_{\rm 2pF}(\bar{R},a,r)-\rho(r)\right|.
\end{align}
for the radius $\bar{R}$ and diffuseness $a$ parameters.
$\rho_0$ is determined by the normalization condition.
Note that the obtained $a$ parameter can be 
extracted accurately from the proton-nucleus elastic scattering cross section
measurement up to the first diffraction peak~\cite{Hatakeyama18}.
Many examples ~\cite{Hatakeyama18,Choudhary20,Choudhary21,Tanaka20, Horiuchi21b,Horiuchi22,Horiuchi22b,Horiuchi23b} showed that the nuclear diffuseness
can be used to deduce the spectroscopic properties of various nuclear systems.

The calculated diffuseness parameters 
are 0.501, 0.544, 0.507, and 0.585 fm for
the $j$-$j$, SU(3), $5\alpha$, and O$\alpha$ configurations,
respectively. We note that the $a$ value becomes
large when the nodal low angular momentum states,
i.e., $1s$ orbit, are occupied~\cite{Horiuchi21b},
while it becomes small in the bubble nuclei~\cite{Choudhary20}. 
The calculated diffuseness clearly reflects
the information on the nuclear surface
showing large $a$ values for the SU(3) and O$\alpha$ configurations,
where the occupation of the $s$ orbit is significant.

\subsection{Ground-state structure of $^{20}$Ne}

\begin{figure}[ht]
  \begin{center}
     \epsfig{file=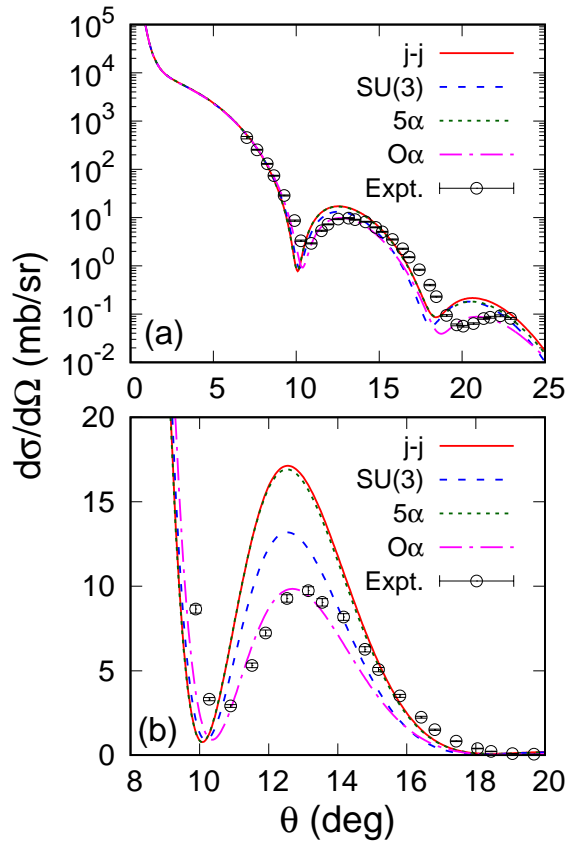, scale=1.4}
  \caption{Proton-nucleus differential elastic scattering cross sections
    for $^{20}{\rm Ne}$ at incident energies of 800~MeV
    as a function of scattering angles in (a) logarithmic and (b) linear scales.
    The experimental data are taken from Ref.~\cite{Blanpied88}.}
    \label{dcs.fig}
  \end{center}
\end{figure}

Those four different density profiles can be distinguished
by the proton-nucleus elastic scattering~\cite{Horiuchi22c,Horiuchi23}.
Figure~\ref{dcs.fig} compares
the differential cross sections for the proton-$^{20}$Ne elastic scattering.
Incident proton energy is chosen to 800~MeV, where
the experimental data are available.
The cross sections in logarithmic and linear scales are plotted.
As we see in the figure,
the differences between those four configurations are apparent.
The bicluster configuration is best for reproducing the data,
explaining the available data up to the second peak position.

The $5\alpha$ and $j$-$j$ coupling shell model
configurations significantly overestimate the experimental data. 
The sharper the nuclear surface, the larger the cross sections
at the first peak position become~\cite{Hatakeyama18}.
In fact, these configurations give significantly smaller $a$ values
compared to the others as were given in the previous subsection.
For the $5\alpha$ configuration,
because ``sharp'' five $\alpha$ particles are located near
the nuclear surface, the cross sections
near the first diffraction peak are enhanced
compared to the O$\alpha$ configuration.
As the $j$-$j$ type has no $1s$ components,
the nuclear surface becomes sharp
with the occupancy of the higher angular momentum state,
i.e., $0d_{5/2}$ orbit~\cite{Choudhary20}.
The SU(3) shell model configuration shows intermediate
between the bubble and bicluster density profiles.
This can be explained by investigating the degree
of nuclear deformation.

As a measure of the nuclear quadrupole deformation,
we calculate the reduced quadrupole transition probabilities
from $J^\pi=2^+$ to the ground $0^+$ states [$B(E2\downarrow)$]
by using parity and angular momentum projected intrinsic wave functions $\Phi$
for each configuration as in Ref.~\cite{Kanada95}. 
The calculated $B(E2)$ values are 6.42, 18.8, 51.0, and 41.9 $e^2{\rm fm}^4$
for $j$-$j$, SU(3), $5\alpha$, and O$\alpha$ types, respectively.
As expected, the $j$-$j$ coupling configuration gives
the smallest $B(E2)$ value.
These for the cluster type configurations are
large and comparable to the experimental data,
56.0$\pm$8.0 $e^2$fm$^4$~\cite{Singhal73,Pritychenko16}.
The value for the SU(3) configuration is
approximately half of these clustered configurations.
Though the one-body density of the SU(3) configuration
has a diffused nuclear surface due to the mixture
of the $s$-orbit, the present SU(3) shell model configuration has
a less deformed shape, which is not enough
to explain the elastic scattering cross sections
at the first peak position.

\begin{figure}[ht]
  \begin{center}
    \epsfig{file=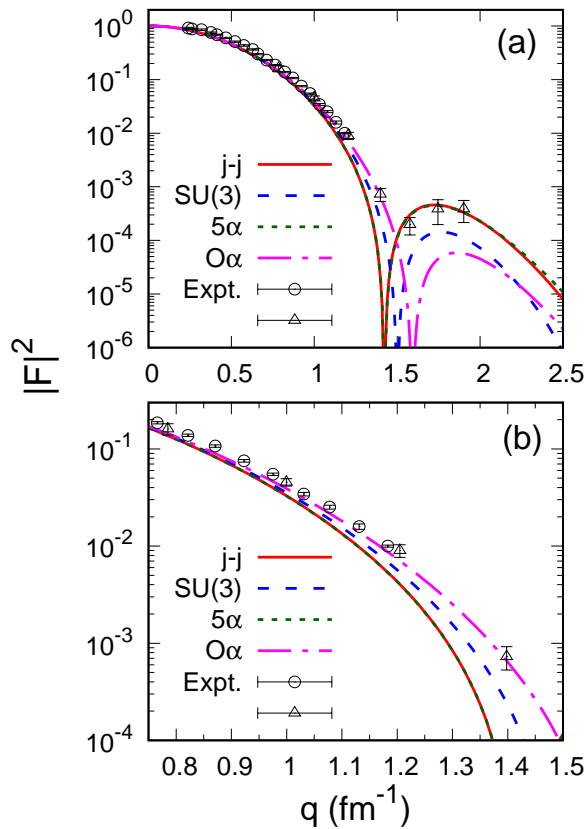, scale=1.4}
  \caption{Squared elastic charge form factors of $^{20}$Ne.
    as a function of the momentum transfer
    in ranges of (a) [0:2.5] fm$^{-1}$ and [0.75:1.5] fm$^{-1}$.
    The experimental data are taken from Refs.~\cite{Horikawa71,Knight81}.
    }
    \label{form.fig}
  \end{center}
\end{figure}

The proton-elastic scattering has
of particular sensitivity to the density profiles
near the nuclear surface~\cite{Hatakeyama18}.
For a more detailed understanding of
the structure of $^{20}$Ne, we evaluate the elastic charge form factor,
which has traditionally been investigated as
a direct observable of the nuclear charge distribution using
electron scattering~\cite{Hofstadter56}.
Since the electron-nucleus interaction is weaker than
that of the proton-nucleus one,
the electron scattering can probe more
internal regions of the charge density distribution
than the proton scattering. The elastic charge form factor is evaluated by
the Fourier transform of the density distribution with
the convolution of the finite proton charge as~\cite{Kamimura81, Horiuchi23}
\begin{align}
  |F(q)|^2=\left|\frac{1}{A}\int_0^{\infty}dr\, \rho(r)j_0(qr) r^2\right|^2
  \exp\left(-\frac{1}{2}a_p^2q^2\right)
  \label{form.eq}
\end{align}
with $a_p^2=0.514$ fm$^2$, which reproduces
the Wm's charge radius of a proton, $0.878$ fm~\cite{Angeli13}.

Figure~\ref{form.fig} (a) displays the square of the elastic charge form factors
of $^{20}$Ne as a function of the momentum transfer.
The Coulomb distortion effect on the momentum transfer is taken into account
in the experimental data~\cite{Horikawa71,Knight81}.
For the sake of visibility, we also draw it for low-momentum
transfer region in panel (b).
We see that the O$\alpha$ configuration perfectly reproduces
the form factors up to $q\approx 1.4$ fm$^{-1}$
but it fails to reproduce the data at higher $q$,
while the $j$-$j$ and $5\alpha$ configurations reproduce
the high $q$ data.
This suggests that the wave function of $^{20}$Ne
is a bicluster-like structure in the surface region but
it could be more depressed in the internal region.
This fact does not contradict with
the conclusion in the proton-elastic scattering
because a proton probe can only be sensitive to the surface region
around the radius of the half density~\cite{Hatakeyama18}
and the internal region cannot directly be detected~\cite{Choudhary20}.
It leads to the interpretation that
the shell-model-like configuration is favored in the internal region
of the density distribution,
while its surface region is dominated by the bicluster configuration.
It is interesting to study the wave function
that includes configuration mixing
for unveiling detailed structure of $^{20}$Ne.
Also, we remark that the possibility of $\alpha$ cluster breaking in
the internal region owing to the spin-orbit interaction
was pointed out in Ref.~\cite{AQCM04}, and
recent $\alpha$ knockout reaction analysis showed that
the spectroscopic factor of $^{16}{\rm O}+\alpha$ configuration
is 0.26 in the ground state using the antisymmetrized molecular
dynamics wave function~\cite{Yoshida19},
which is smaller than the cluster model prediction, 0.3-0.4~\cite{Nemoto75}.

\begin{figure}[ht]
\begin{center}
  \epsfig{file=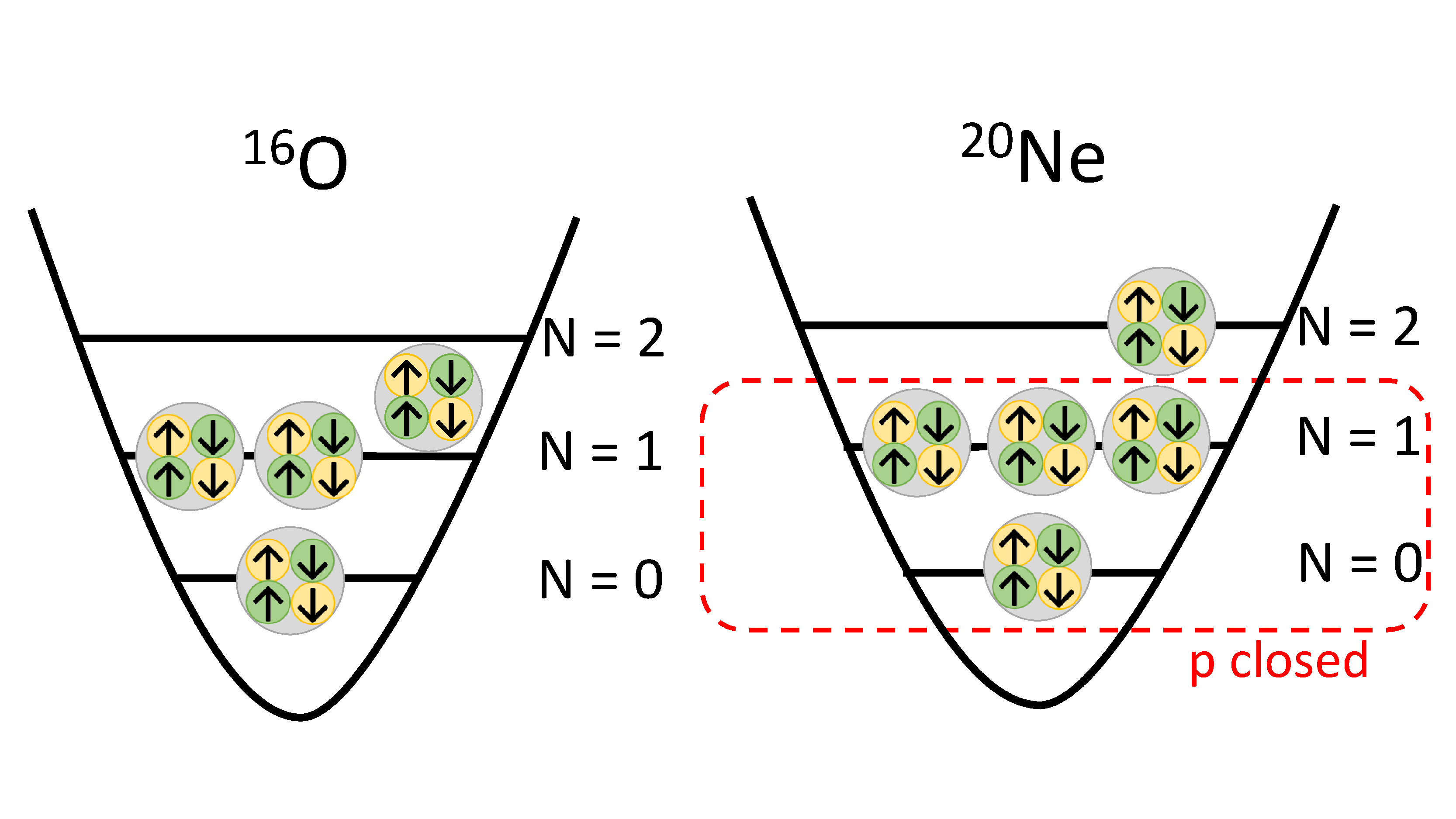, width=\linewidth}
  \caption{Schematic picture of configurations of
    tetrahedron $^{16}$O and bicluster $^{20}$Ne configurations.}
    \label{bicluster.fig}
  \end{center}
  \end{figure}

In the present analysis, we assume that the ``$^{16}$O cluster'' in $^{20}$Ne
has $p$-shell closed configuration.
It seems to contradict with the fact that $^{16}$O wave function in vacuum
has the 4$\alpha$ tetrahedron configuration~\cite{Horiuchi23}.
This can be explained as follows:
The $4\alpha$ tetrahedron configuration is not an ideal $p$-shell closed
state but includes the mixture of $sd$ shell but $1s$ orbit
deficient~\cite{Horiuchi23}.
We calculate the squared overlap between the tetrahedron $4\alpha$ and $p$-shell configurations of $^{16}$O and find
  that it is small, 0.024.
Note that this value can be understood
if the squared overlap between the Gaussian wave
packets are 80\% and evaluate $(0.8)^{16}=0.03$.
The overlap between the many-body wave functions
  is very sensitive to the shapes of the single-particle wave functions.
  The reason of this small squared overlap is partly because 
  large intercluster distances $d$ in the tetrahedron configuration
  and partly the difference in the $\nu$ parameters.
  For example, the squared overlap value is recovered
  to 0.12 ($\sim 88$\% for the average squared overlap of the Gaussian wave packets) when
  we take $d\to 0$ for the tetrahedron $4\alpha$ configuration.
In $^{20}$Ne, an additional $\alpha$ particle fills
completely the $p$-shell and partially the $sd$-shell,
and thus the $p$-shell closed 16 nucleons
plus $\alpha$ cluster structure is realized.
A schematic picture is drawn in Fig.~\ref{bicluster.fig}.
  From this interpretation, the O$\alpha$ configuration
  should include a certain amount of
  the SU(3) configuration. In fact, the squared overlap between
  these configurations is large, 0.39.

\section{Conclusion}
\label{conclusion.sec}

To explore the most probable configuration
of the ground state of $^{20}$Ne,
we have made a comprehensive analysis by using various configurations
generated from the antisymmetrized quasicluster model (AQCM).
We have examined four AQCM configurations of
the (a) $j$-$j$ coupling and (b) SU(3) shell model,
(c) $5\alpha$ and (d) $^{16}{\rm O}+\alpha$-like cluster configurations.
With these that reproduce the experimental charge radius data,
we have calculated one-body density distributions
and evaluated physical observables that
directly reflect the characteristics of the density profiles.

We find that the characteristics of these four configurations
are imprinted on the density profile near the nuclear surface,
which can clearly be distinguished by comparing
theoretical and experimental proton-nucleus elastic scattering cross sections.
We conclude that the ground-state structure of $^{20}$Ne
includes a significant amount of a 16+4 nucleon bicluster structure.
We note, however, that
the 16 nucleons does not necessarily mean $^{16}$O in vacuum,
i.e., $4\alpha$ tetrahedron, but can be interpreted as
the $p$-shell closed 16 nucleon configuration.

Also, the elastic charge form factor at high momentum region suggests
that the breaking of the $\alpha$ cluster around $^{16}$O may happen
in the internal region of $^{20}$Ne.
The competition of $j$-$j$ coupling shell model and $\alpha$
cluster structure and the relation
to the spin-orbit interaction is an intriguing subject.

As an extension of the present study,
exploring various cluster structure along heavier $N=Z$ nuclei
such as $^{24}$Mg, $^{28}$Si, $^{32}$S, and $^{36}$Ar is interesting
and will be reported elsewhere soon.

\section*{acknowledgments}

This work was in part supported by JSPS KAKENHI Grants
Nos.\ 18K03635, 22H01214, and 22K03618.

\end{document}